\documentclass[showpacs,floatfix,eqsecnum,preprintnumbers,aps,nofootinbib]{revtex4}

\usepackage{epsfig}
\usepackage{citesort}

\def  \vtbvtd     {V_{tb} V_{td}^*}
\def  \vcbvcd     {V_{cb} V_{cd}^*}
\def  \vubvud     {V_{ub} V_{ud}^*}
\def  \vtbvts     {V_{tb} V_{ts}^*}
\def  \vcbvcs     {V_{cb} V_{cs}^*}
\def  \vubvus     {V_{ub} V_{us}^*}
\def  \lamu       {\lambda_u}
\def  \la         {\langle }
\def  \ra         {\rangle }

\def  \matfac     {\frac{G_F \alpha}{\sqrt{2} \pi} \vtbvtd} 

\def  \tgb        { \tan\beta }

\def  \btopill    {B \to \pi \ell^+ \ell^-}
\def  \btorholl   {B \to \rho \ell^+ \ell^-}

\def  \btobpill   {B \to \bar{\pi} \ell^+ \ell^-}
\def  \btobrholl  {B \to \bar{\rho} \ell^+ \ell^-}
\def  \bbtopill   {\bar{B} \to \pi \ell^+ \ell^-}
\def  \bbtorholl  {\bar{B} \to \rho \ell^+ \ell^-}

\def  \btopitt    {B \to \pi \tau^+ \tau^-}
\def  \btorhott   {B \to \rho \tau^+ \tau^-} 
\def  \bbtopitt   {\bar{B} \to \pi \tau^+ \tau^-}
\def  \btobpitt   {B \to \bar{\pi} \tau^+ \tau^-}
\def  \bbtorhott  {\bar{B} \to \rho \tau^+ \tau^-}
\def  \btobrhott  {B \to \bar{\rho} \tau^+ \tau^-} 

\def  \btopirholl {B \to (\pi , \rho) \ell^+ \ell^-}

\def  \btoxdll    {B \to X_d \ell^+ \ell^-}
\def  \btoxsdll   {B \to X_{s,d} \ell^+ \ell^-}
\def  \btokksll   {B \to K(K^*) \ell^+ \ell^-}
\def  \btosll     {b \to s \ell^+ \ell^-}
\def  \btodll     {b \to d \ell^+ \ell^-}
\def  \btosdll    {b \to s(d) \ell^+ \ell^-}

\def  \csev       { C_7^{eff} }
\def  \cn         { C_9^{eff} } 
\def  \ct         { C_{10} }
\def  \cqo        { C_{Q_1} }
\def  \cqt        { C_{Q_2} }

\def  \bd         { \bar{d} }
\def  \bl         { \bar{\ell} }
\def  \mb         { m_b }
\def  \mcb        { m_B }
\def  \mpi        { m_\pi } 
\def  \mrh        { m_\rho }
\def  \ml         { m_\ell }

\def  \ppi        { p_\pi}
\def  \pcb        { p_B}

\def  \sh         { \hat{s} }
\def  \mlh        { \hat{m}_\ell }
\def  \mlhs       { \hat{m}_\ell^2 }
\def  \mpihs      { \hat{m}_\pi^2 }
\def  \mrhohs     { \hat{m}_\rho^2 }
\def  \faco       { \sqrt{1 - \frac{4 \mlhs}{\sh}} } 

\def  \dgdsdz     { \frac{d \Gamma}{d \sh dz}} 
\def  \pv         { {\bf p}_- }
\def  \qv         { {\bf q} }
\def  \ev#1       { {\bf e}_#1 }

\def  \pl         { P_L }
\def  \pr         { P_R }
\def  \gf         { \gamma_5 }
\def  \gm         { \gamma_\mu }

\def  \gom        { \gamma^\mu }
\def  \sigmn      { \sigma_{\mu \nu} }
\def  \qm         { q_\mu }
\def  \qon        { q^\nu }
\def  \qs         {q^2}

\def  \fz        { F_0(\qs) }
\def  \fo        { F_1(\qs) }
\def  \ft        { F_T(\qs) } 
\def  \ftt       { \tilde{F}_T (\qs) } 
\def  \lampi     { \lambda(1,\sh,\mpihs) }

\def  \prh       { p_\rho }
\def  \amnab     { \epsilon_{\mu \nu \alpha \beta} } 
\def  \lamrho    { \lambda(1,\sh,\mrhohs)} 
\def  \bturho    { \bigtriangleup_\rho }

\def \beq       {\begin{equation}}
\def \eeq       {\end{equation}}
\def \beqa      {\begin{eqnarray}}
\def \eeqa      {\end{eqnarray}}
\def \bfig      {\begin{figure}}
\def \efig      {\end{figure}}
\def \bcen      {\begin{center}}
\def \ecen      {\end{center}}

\def \ie        {, i.e. ,}

\def \etal      { et. al. }

\def \prd#1#2#3       {Phys. \ Rev. {\bf D #1}, {#2} (#3)}
\def \prl#1#2#3       {Phys. \ Rev. \ Lett. {\bf #1}, {#2} (#3)}
\def \nuclphysb#1#2#3 {Nucl. \ Phys. {\bf B #1}, {#2} (#3)}
\def \plb#1#2#3       {Phys. \ Lett. {\bf B #1}, {#2} (#3)}
\def \physrep#1#2#3   {Phys. \ Rep {\bf #1}, {#2} (#3)}
\def \zphysc#1#2#3    {Z. \ Phys. {\bf C #1}, {#2} (#3)}

\begin{document}

%%%%%%%%%%%%%%%%%%%%%%%%%%%%%%%%%%%%%%%%%%%%%%%%%%%%%%%
%%  Title page 

%% Preprint number 
\preprint{hep-ph/0206128}

\title{SUSY effects on the exclusive semi-leptonic decays $\btopitt$
and $\btorhott$ } 

\author{S.Rai Choudhury}
 \email{src@ducos.ernet.in}
\author{Naveen Gaur}
 \email{naveen@physics.du.ac.in}
 \homepage{http://physics.du.ac.in/~naveen}

\affiliation{Department of Physics \& Astrophysics, \\ 
  University of Delhi,  \\
  Delhi - 110 007, India} 
\date{\today}


%%%%%%%%%%%%%%%%%%%%%%%%%%%%%%
%  Abstract

\begin{abstract}
Semi-leptonic decays of B-meson are known to be very sensitive to the
any new physics effects. Amongst various possibilities the transition
at the quark level $\btodll$ is more suited than $\btosll$ for purpose
of study of CP violation. Here in this work we will discuss the
effects of Supersymmetry on the various experimentally measurable
quantities like decay rate, Forward backward asymmetry, various
polarization asymmetries and CP violation asymmetries in exclusive
channels $\btopill$ and $\btorholl$. We will be going to focus mainly
on the Neutral Higgs boson effects on these measurements,
with a view of eliciting information about possible CP-violating as
well as non-hermitian terms in the effective Hamiltonian.  
\end{abstract}

\maketitle

%%%%%%%%%%%%%%%%%%%%%%%%%%%%%%%%%%%%%%%%%%%%%%%%%%%%%%%%
%%%%%%%%%%%%  Section

\section{\label{section:1}Introduction}

The rare B meson decays induced by Flavor changing neutral current
(FCNC) $b \to s(d)$ transition offer a deeper probe for weak
interaction sector of the Standard model (SM) as they go through
second order in weak interaction . These decays can give information
regarding the fundamental parameters like Cabbibo-Kobayashi-Maskawa
(CKM) factors, leptonic decay constants etc. These decays can also be
very useful in testing out the various new physics scenarios like
2HDM models \cite{Iltan:2000iw,Skiba:1993mg}, MSSM
\cite{Choudhury:1999ze,Xiong:2001up,RaiChoudhury:1999qb,goto1,Cho:1996we}. 
Among the weak decays of B-meson the leptonic and semi-leptonic decays
are very useful because of their relative cleanness. Among these too
the exclusive decays lately have received special attention
\cite{Demir:2002a,Iltan:1998a}. For the semi-leptonic and
leptonic decays like $\btoxsdll, \btokksll, \btopill, \btorholl$
etc. the basic quark level process is $\btosdll$
\cite{Ali:1997vt}. The basic quark level process $\btosdll$ occurs
through intermediate t, c or u quark. These processes can be described
in terms of effective Hamiltonian which contains the information about
short and long distance effects. For the quark level process $\btosll$
the various contributions due to intermediate t, c and u quarks enters
in matrix elements with factors $\vtbvts ,~ \vcbvcs$ and
$\vubvus$. Off these three factors the third one is extremely small as
compared to the first two. The unitarity relationship of CKM matrix
becomes (approximately) $\vtbvts + \vcbvcs \approx 0$, so that the
second factor can be written effectively as the negative of the first
one. The Hamiltonian (effective) for $\btosll$ transition thus
involves essentially only one independent CKM factor $\vtbvts$ and
hence the process $\btosll$ is not sensitive to CKM phases within SM.
\cite{Du:1995ez}.    

\par For the transition $\btodll$ the CKM factors involved $\vtbvtd,
~\vcbvcd$ and $\vubvud$ are comparable in magnitude and so the
crossection for the processes having the quark 
level process $\btodll$ can have significant interference terms
between them, and this could open up the possibility of observing the
complex CKM factors from the interference terms. In the semi-leptonic
decays (having lepton pair in final state ) one can discuss several
other kinematical variables associated with final state leptons, like
lepton pair Forward Backward (FB) asymmetry and various polarization
asymmetries. SUSY effects on the FB asymmetries in various exclusive
decay modes of  B-meson, like $B \to (\pi,K) \ell^+ \ell^- $
\cite{Demir:2002a}, $B \to (\pi, \rho) \ell^+ \ell^-$
\cite{Demir:2002a,Iltan:1998a}, $B \to \ell^+ \ell^- \gamma$
\cite{RaiChoudhury:2002i}  have been extensively studied. Sometime
back as pointed out by one of us \cite{RaiChoudhury:1997a} for
inclusive decay mode $\btoxdll$ and by Kr\"{u}ger and Sehgal
\cite{Kruger:1997jk} for exclusive mode $\btopirholl$ that along with
CP violation from partial width asymmetry one can look for CP
violation in FB asymmetry also, we will be going to explore this
possibility also.   

\par Along with FB asymmetry associated with final state leptons one
can also discuss the three polarization asymmetries (longitudinal,
normal and transverse) associated with final state leptons in various
semi-leptonic decays. The importance of polarization asymmetries
associated with final state leptons in various inclusive and exclusive
semi-leptonic decay modes have been extensively discussed in many
works
\cite{RaiChoudhury:1999qb,Kruger:1996cv,Fukae:1999ww,RaiChoudhury:2002i}
. In this communication, we study the three possible polarization
asymmetries with the idea of exploring CP asymmetries as well as the
{\sl non hermiticity} of the effective Hamiltonian\footnote{which
arises out of the quark loops} through their measurement. 

\par The paper is organized as follows : In section \ref{section:2} we
will be going to present the general formalism where we will be going
to write the general effective Hamiltonian and will be going to
present our definitions of the FB asymmetry and polarization
asymmetries. Section \ref{section:3} will be devoted to the decay mode
$\btopill$ where we will discuss all the kinamatical variables
associated with this decay mode. In Section \ref{section:4} we will
discuss the decay mode $\btorholl$. Finally we will conclude with
results and discussion in section \ref{section:5}. 

%%%%%%%%%%%%%%%%%%%%%%%%%%%%%%%%%%%%%%%%%%%%%%%%%%%%%%%%
%%%%%%%%  Section : 2

\section{\label{section:2} General Formalism}

The QCD corrected effective Hamiltonian for the decay $\btodll$ in
general SUSY model can be written as
\cite{Xiong:2001up} :
\beq
{\cal H}_{eff} 
   ~=~ \frac{4 G_F \alpha }{\sqrt{2} \pi} \vtbvtd
       \Bigg[ \sum_{i = 1}^{10} C_i O_i 
          ~+~ \sum_{i=1}^{10} C_{Q_i} Q_i 
      ~-~ \lamu \left\{ ~C_1 [ O_1^u - O_1 ] ~+~ C_2 [ O_2^u - O_2 ] ~
         \right\} 
       \Bigg] ,
\label{sec2:eq:1}
\eeq
where we have used the unitarity of the CKM matrix $\vtbvtd ~+~ \vubvud
\approx - \vcbvcd$, and $\lamu ~=~ \vubvud /\vtbvtd$. Here $O_1$ and
$O_2$ are the current current operators, $O_3, \dots, O_6$ are called
QCD penguin operators and $O_9$ and $O_{10}$ are semileptonic
electroweak penguin operators \cite{Ali:1997vt}. The new operators
$Q_i \rm (i = 1,\dots,10)$ arises due to NHB (Neutral Higgs Boson)
exchange diagrams \cite{Skiba:1993mg,Xiong:2001up}. 
%% New addition
In addition to the short distance corrections included in the Wilson
coefficients, there are some long distance effects also, associated
with real $c \bar{c}$ resonances in the intermediate states. This is
taken into account by using the prescription given in
\cite{Long-Distance} namely by using the Breit-Wigner form of the
resonances that adds to $\cn$ :
\beq 
C_9^{res} = \frac{- 3 \pi}{\alpha^2} \kappa_v
\sum_{V = J/\psi,\psi',..}
\frac{M_V Br(V \to \ell^+ \ell^-)\Gamma_{\rm total}^V}{(s - M_V^2)
+ i \Gamma_{\rm total}^V M_V} ,
\label{sec2:eq:1a}
\eeq
as we are taking final leptons to be $\tau$ so only five resonances of
the $c \bar{c}$ will contribute. The phenomenological factor
$\kappa_v$ is taken to be $2.3$ for numerical calculations
\cite{Kruger:1996cv} 
%%%%%%%%%
In this work we
will be going to use the Wolfenstein parameterisation
\cite{Wolfenstein:1983yz} of CKM matrix with four real  parameters
$\lambda, A, \rho ~{\rm and}~ \eta$ where $\eta$ is the measure of CP
violation. In terms of these parameters we can write $\lamu$ as :   
\beq
\lamu ~=~ \frac{\rho (1 - \rho) - \eta^2}{(1 - \rho)^2 + \eta^2} 
~-~ i \frac{\eta}{(1 - \rho)^2 + \eta^2} ~+~ O(\lambda^2)
\label{sec2:eq:2} , 
\eeq
From the relevant part of the above effective Hamiltonian
eqn.(\ref{sec2:eq:1}) we can write the QCD corrected matrix element as
: 
\beqa
{\cal M} 
 &=& \frac{G_F \alpha}{\sqrt{2} \pi} \vtbvtd 
   \left\{ ~ - 2 ~\csev ~\frac{\mb}{\qs} ~( \bd i \sigmn \qon \pr b )~
         ( \bl \gom \ell) ~+~ \cn ~ (\bd \gm \pl b)~ (\bl \gom \ell) 
      ~+~ \ct ~(\bd \gm \pl b )~ (\bl \gom \ell)  
   \right.  \nonumber  \\ 
 && \left. + ~\cqo ~(\bd \pr b)~(\bl \ell) 
    ~+~ \cqt ~(\bd \pr b)~ (\bl \gf \ell) ~  \right\} , 
\label{sec2:eq:3}
\eeqa
where q is the momentum transfer and $P_{L,R} ~=~ (1 \mp \gf)/2$, we
have neglected mass of d quark. The wilson coefficients $\csev$ and
$\ct$ are given in many works
\cite{Xiong:2001up,Cho:1996we,Grinstein:1989me} and the other Wilsons 
$\cqo$ and $\cqt$ are given in
\cite{Choudhury:1999ze,Xiong:2001up}. The
definition of $\cn$ is given in
\cite{RaiChoudhury:1997a,Kruger:1996dt, Iltan:1998a}.

\par The decay rate (for any general three body decay process $B \to P
\ell^+ \ell^-$) can be evaluated by doing the phase space
integration. On doing the phase space integration we can get 
\beq
\frac{d \Gamma(B \to P \ell^+ \ell^-)}{d \sh dx} 
~=~ \frac{\mcb}{2^9 \pi^3} ~\lambda^{1/2}(1,\sh,\hat{m}_p^2) 
~\sqrt{1 - \frac{4 \mlhs}{\sh}} ~|{\cal M}|^2 , 
\label{sec2:eq:4}
\eeq
where $\sh = s/\mcb^2,~\mlh = m_\ell/\mcb, ~\hat{m}_p~=~ m_p/\mcb $
are the dimensionless quantities. $\lambda(a,b,c) ~=~ a^2 + b^2 + c^2
- 2 a b - 2 a c - 2 b c$ . s is the CM energy of $\ell^+ \ell^-$
system, $m_P$ is the mass of particle labeled as P and $z ~=~
Cos\theta$ where $\theta$ is the angle between $\ell^-$ and $B$
three momenta in CM frame of $\ell^+ \ell^-$. $|{\cal M}|^2$ is the 
matrix element square of the process under consideration. 

\par From above expression one can get decayrate, FB asymmetry
\cite{Long-Distance} . Decay rate is simply the integration of
eqn(\ref{sec2:eq:4}) over the angle z. The definition of the FB
asymmetry is \cite{Long-Distance}: 
\beq
A_{FB} ~=~  
   \frac{ \int_0^1 dz  \dgdsdz - \int_{-1}^0 dz
       \dgdsdz }{\int_0^1 dz  \dgdsdz + \int_{-1}^0
       dz \dgdsdz } , 
\label{sec2:eq:5}
\eeq

For defining the polarization asymmetries we define the orthogonal
unit vectors, $S$ in the rest frame of $\ell^-$ for the polarization
of lepton $\ell^-$ \cite{Kruger:1996cv,RaiChoudhury:1999qb} to the
longitudinal direction ($L$), the normal direction ($N$) and the
transverse direction ($T$).   
\beqa
S^\mu_L &\equiv& (0, \ev{L}) ~=~ (0, \frac{\pv}{|\pv|})  \nonumber \\
S^\mu_N &\equiv& (0, \ev{N}) ~=~ (0, \frac{\qv \times \pv}{|\qv \times
                     \pv})         \nonumber   \\
S^\mu_T &\equiv& (0, \ev{T}) ~=~ (0, \ev{N} \times \ev{L}) 
\label{sec2:eq:6}
\eeqa
where $\pv $ and $\qv $ are the three momenta of $\ell^-$ and photon
in the center-of-mass (CM) frame of $\ell^- \ell^+$ system. Now
boosting all the three vectors given in eqn.(\ref{sec2:eq:6}), only 
longitudinal vector will get boosted and rest two (normal and
transverse) will remain the same. Longitudinal vector after boost
becomes 
\beq
S^\mu_L ~=~ ( \frac{|\pv|}{m_\ell}, \frac{E_- \pv}{m_\ell |\pv|} )
\label{sec2:eq:7}
\eeq
Now we can calculate the polarization asymmetries by using the spin
projectors for $\ell^-$ as ${1 \over 2} ( 1 + \gamma_5 \not S)$. The
lepton polarization asymmetries are defined as :
\beq
P_x(\sh) ~~~\equiv~~~
               \frac{ \frac{d \Gamma(S_x)}{d \sh}  ~-~ \frac{d \Gamma(-
                      S_x)}{d \sh}}{\frac{d \Gamma(S_x)}{d \sh}  
                      ~+~ \frac{d \Gamma(- S_x)}{d \sh}}
\label{sec2:eq:8}
\eeq
with $x = L,~T,~N$ respectively for longitudinal, transverse and
normal polarization asymmetry.

\par Now ready with the terminology and definitions we will move to
the calculations of the various measurable quantities that we
mentioned before.

%%%%%%%%%%%%%%%%%%%%%%%%%%%%%%%%%%%%%%%%%%%%%%%%%%%%%%%%%
%%%%%%%%  Section : 3

\section{\label{section:3} \boldmath $\btopill$ decay mode}

In this section we calculate the Branching ratio, FB asymmetry and the
polarization asymmetries associated with the inclusive decay mode
$\btopill$. Using the definition of the form factors (
eqn(\ref{appendixa:eq:1})-(\ref{appendixa:eq:3}) ) we can write down
the matrix element for $\bar{B} \to \pi$ \footnote{$\bar{B}$
actually is $B^+$} transition as \footnote{in writting this we have
used $\bl \not q ~\ell ~=~ 0 \ ,\ \bl \not q \gf ~\ell ~=~ 2 m_\ell \bl
\gf \ell$} :
\beq
{\cal M}^{\bar{B} \to \pi} ~=~ \matfac \left\{ A (\pcb)_\mu (\bl \gm
\ell) ~+~ B (\pcb)_\mu  (\bl \gm \gf \ell) ~+~ C (\bl \gf \ell) ~+~ D
(\bl \ell) \right\} 
\label{section3:eq:1}
\eeq
with
\beqa
A &=& \cn \fo ~-~ 2 \csev \ftt                        \\
\label{section3:eq:2}
B &=& \ct \fo                                         \\
\label{section3:eq:3}
C &=& \ml \ct \left\{ - \fo ~+~ \frac{(\mcb^2 - \mpi^2)}{\qs} (\fz -
\fo) \right\} ~+~ \frac{(\mcb^2 - \mpi^2)}{2 \mb} \fz \cqt   \\ 
\label{section3:eq:4}
D &=&  \frac{(\mcb^2 - \mpi^2)}{2 \mb} \fz \cqo    
\label{section3:eq:5}
\eeqa
From the above expression of the matrix element
(eqn.(\ref{section3:eq:1})) we can get the analytical expression of
the decay rate as 
\beq
\frac{d \Gamma(\bbtopill)}{d\sh} = \frac{G_F^2 \mcb^5 \alpha^2}{3
\times 2^9 \pi^5}~ |\vtbvtd|^2  \lambda^{1/2}(1,\sh,\mpihs) ~ \faco ~
\Sigma_\pi  
\label{section3:eq:6}
\eeq
with 
\beqa
\Sigma_\pi &=&  \lampi \left(1 + \frac{2 \mlhs}{\sh} \right) |A|^2 ~+~
\left[ \lampi \left(1 + \frac{2 \mlhs}{\sh} \right) + 24 \mlhs \right]
|B|^2 ~+~ 6 \frac{\sh}{\mcb^2} |C|^2       \nonumber      \\
&& ~+~ 6 \frac{\sh}{\mcb^2} 
\left(1 - \frac{4 \mlhs}{\sh} \right) |D|^2 ~+~ 12 ~ \frac{\mlh}{\mcb}
~(1 + \sh - \mpihs) ~ Re(C^* B) 
\label{section3:eq:7}
\eeqa

%%%%%%%%%%%%%%%%%%%%%%%%%%%%%%%%%%%%%%%%%%%%%%%%%%%%%%%%%%%
%%  Figure : 1
%\vskip 0.9cm
\bfig[t]
\epsfig{file=btopidrs.eps,height=7.5cm}
\caption{Branching ratio for $\bbtopitt$. Other parameters are :
mSUGRA $m = 200 ~,~ M = 450 ~,~ A = 0 ~,~ \tgb = 35 ~,~ sgn(\mu) $ is
taken to be +ve for rSUGRA $m_A = 306$. All masses are in GeV }
\label{fig:1}
\efig
%%%%%%%%%%%%%%%%%%%%%%%%%%%%%%%%%%%%%%%%%%%%%%%%%%%%%%%%%%%

The expression of FB asymmetry is :
\beq
A_{FB}(\bbtopill) ~=~ - 6 ~ \mlh ~ \lambda^{1/2}(1,\sh,\mpihs) \faco
 ~\frac{Re(A D^*)}{(\mcb \Sigma_\pi )}
\label{section3:eq:8}
\eeq
As we can see from above expression that FB asymmetry is proportinal
to the new interactions \ie NHB contributions. This is a point which
has also been noted in some earlier works
\cite{Demir:2002a}. 

%%%%%%%%%%%%%%%%%%%%%%%%%%%%%%%%%%%%%%%%%%%%%%%%%%%%%%%%%%%
%%  Figure : 2
\bfig[!bp]
\vskip 0.8cm
\epsfig{file=btopifb.eps,height=7.5cm}
\caption{FB asymmetry for $\bbtopitt$. Other parameters are :
mSUGRA $m = 200 ~,~ M = 450 ~,~ A = 0 ~,~ \tgb = 35 ~,~ sgn(\mu) $ is
taken to be +ve for rSUGRA $m_A = 306$. All masses are in GeV }
\label{fig:2}
\efig
%%%%%%%%%%%%%%%%%%%%%%%%%%%%%%%%%%%%%%%%%%%%%%%%%%%%%%%%%%%

We divide this section in two subsections in the first one we will
discuss the CP violation in $\btopill$ and in next subsection we will
discuss the polarization asymmetries associated with final state
leptons.

\subsection{\label{section3:subsec:1} CP violation}

Firstly we define CP-violating partial width asymmetry between B and
$\bar{B}$ decay. This is defined as :
\beq
A_{CP}(\sh) ~=~ \frac{\frac{d \Gamma}{d\sh} - \frac{d
\bar{\Gamma}}{d\sh}}{\frac{d \Gamma}{d\sh} + \frac{d
\bar{\Gamma}}{d\sh}} 
\label{section2:eq:9}
\eeq
where
\beq
\frac{d\Gamma}{d\sh} ~=~ \frac{d\Gamma(\bbtopill)}{d\sh} ~~,~~
\frac{d\bar{\Gamma}}{d\sh} ~=~ \frac{d\Gamma(\btobpill)}{d\sh}
\label{section2:eq:10}
\eeq
In going from $\Gamma$ to $\bar{\Gamma}$ the only change we have to
make is in the expression of $\cn$. Defining $\cn$ as :
\beq
\cn ~=~ \xi_1 ~+~ \lamu ~\xi_2
\label{section3:eq:11}
\eeq
where $\xi_1 ~,~ \xi_2$ and $\lamu$ all are complex. In going from
$\bbtopill$ to $\btobpill$,  $\cn$ becomes :
\beq
\cn ~=~ \xi_1 ~+~ \lamu^* \xi_2
\label{section3:eq:12}
\eeq
with this change one can get the expression for $d\bar{\Gamma}/d\sh$
as :
\beq
\frac{d \Gamma(\btobpill)}{d\sh} = \frac{G_F^2 \mcb^5 \alpha^2}{3
\times 2^9 \pi^5}~ |\vtbvtd|^2  \lambda^{1/2}(1,\sh,\mpihs)~ \faco ~ 
\left\{ \Sigma_\pi + 4 ~Im \lamu ~\bigtriangleup_\pi \right\} 
\label{section3:eq:13}
\eeq
with
\beq
\bigtriangleup_\pi ~=~ \left\{ Im (\xi_1^* \xi_2) ~|F_1(s)|^2 - 2
~\csev ~Im\xi_2 ~F_T(s)~ F_1(s)~ \frac{\mb}{(\mcb + \mpi)} \right\}
\lampi ~\left(1 + \frac{2 \mlhs}{\sh}\right)
\label{section3:eq:14}
\eeq
Using eqn.(\ref{section3:eq:6}),(\ref{section3:eq:13}) and
eqn.(\ref{section3:eq:14}) we can get the CP violating partial width
asymmetry as :
\beqa
A_{CP}(\sh) &=& \frac{- ~2 ~Im \lamu ~ \bigtriangleup_\pi}{\Sigma_\pi
~+~ 2~ Im \lamu \bigtriangleup_\pi} 
\label{section3:eq:15}
\eeqa 

%%%%%%%%%%%%%%%%%%%%%%%%%%%%%%%%%%%%%%%%%%%%%%%%%%%%%%%%%%%
%%  Figure : 3
\bfig[ht]
\vskip 0.8cm
\epsfig{file=btopicpdr.eps,height=7.5cm}
\caption{CP violating asymmetry ($A_{CP}$) in  $\bbtopitt$ and
$\btobpitt$. The Wolfenstein parameter we have choosen are : $\rho = -
0.07 ~,~ \eta = 0.34 $. Other parameters are : mSUGRA $m = 200 ~,~ M =
450 ~,~ A = 0 ~,~ \tgb = 35 ~,~ sgn(\mu) $ is taken to be +ve for
rSUGRA $m_A = 306$. All masses are in GeV } 
\label{fig:3}
\efig
%%%%%%%%%%%%%%%%%%%%%%%%%%%%%%%%%%%%%%%%%%%%%%%%%%%%%%%%%%%

As argued in earlier work \cite{RaiChoudhury:1997a} that by measuring
the FB asymmetries of B and $\bar{B}$ one can observe the CP violating
phase of CKM matrix. We will here give as to how much the predictions
of SM would change in SUSY. 

\par In the discussion of CP violation by measuring FB asymmetry its
important to fix the sign of convention. The sign of FB asymmetry for
$B$ and $\bar{B}$ decays are different. In fact under strict CP
conservation :
\beq
A_{FB}(\bar{B}) ~=~  - ~ A_{FB}(B)
\label{section3:eq:16}
\eeq
So under CP conservation the FB asymmetry of $B$ and $\bar{B}$ are
exactly opposite \footnote{we can understand this -ve sign because
under CP conjugation not only $ b \leftrightarrow \bar{b}$ but there
is a transformation in leptons also and $\ell^+ \leftrightarrow
\ell^-$. Since the two leptons are emmitted back to back in cm frame
of dileptons the FB asymmetry defined in terms of negatively charged
lepton ,$\ell^-$ (for both $B$ and $\bar{B}$), this changes sign under
CP conjugation}. So any change in eqn.(\ref{section3:eq:15}) will be
measure of CP violation. We define CP violating parameter of FB
asymmetry as : 
\beq
\delta_{FB} ~=~ A_{FB}(\bbtopill) ~+~ A_{FB}(\btobpill)
\label{section3:eq:17}
\eeq
We can get the expression of $\delta_{FB}$ from the expression of FB
asymmetry eqn.(\ref{section3:eq:8}) as :
\beqa
\delta_{FB} &=&  \frac{12 \mlh \lambda^{1/2}(1,\sh,\mpihs) \faco
}{\mcb \Sigma_\pi (\Sigma_\pi +  4 Im \lamu \bigtriangleup_\pi) } ~ Im
\lamu  \Bigg[ \Sigma_\pi F_1(s) Im \xi_2       \nonumber     \\ 
&& + \bigtriangleup_\pi ( 2 \csev
\tilde{F}_T(s) + F_1(s) + Im \xi_2 Im \lamu - F_1(s) Re\xi_1 - F_1(s)
Re \xi_2 Re \lamu) \Bigg]
\label{section3:eq:18}
\eeqa

%%%%%%%%%%%%%%%%%%%%%%%%%%%%%%%%%%%%%%%%%%%%%%%%%%%%%%%%%%%
%%  Figure : 4
\bfig[ht]
\vskip 0.5cm
\epsfig{file=btopicpfb.eps,height=7.5cm}
\caption{CP violating asymmetry ($\delta_{CP}$) in  $\bbtopitt$ and 
$\btobpitt$. The Wolfenstein parameter we have choosen are : $\rho = -
0.07 ~,~ \eta = 0.34 $. Other parameters are : mSUGRA $m = 200 ~,~ M =
450 ~,~ A = 0 ~,~ \tgb = 35 ~,~ sgn(\mu) $ is taken to be +ve for
rSUGRA $m_A = 306$. All masses are in GeV } 
\label{fig:4}
\efig
%%%%%%%%%%%%%%%%%%%%%%%%%%%%%%%%%%%%%%%%%%%%%%%%%%%%%%%%%%%

\subsection{\label{section3:subsec:2} Polarization asymmetries} 

We can also get the expression of the polarization asymmetries in
$\bbtopill$ transtion using the formalism given in section
\ref{section:2}. The expression of the various polarization
asymmetries of $\ell^-$ are :
\beqa
P_L &=&  0     \label{section3:eq:19}              \\
P_T &=&  0     \label{section3:eq:20}              \\
P_N &=&  \frac{\pi ~D ~\lambda^{1/2}(1,\sh,\mpihs) ~\sqrt{\sh - 4
\mlhs}}{\Sigma_\pi} ~  \Bigg[ Im\xi_1 + Im \lamu Re \xi_2 + Im \xi_2
Re \lamu \Bigg] 
\label{section3:eq:21}
\eeqa

%%%%%%%%%%%%%%%%%%%%%%%%%%%%%%%%%%%%%%%%%%%%%%%%%%%%%%%%%%%
%%  Figure : 5
\bfig[ht]
\vskip 0.5cm
\epsfig{file=btopinorm.eps,height=7.5cm}
\caption{Normal polarization asymmetry ($P_N$) in  $\bbtopitt$ . The
Wolfenstein parameter we have choosen are : $\rho = - 0.07 ~,~ \eta =
0.34 $. Other parameters are : mSUGRA $m = 200 ~,~ M = 450 ~,~ A = 0
~,~ \tgb = 35 ~,~ sgn(\mu) $ is taken to be +ve for rSUGRA $m_A =
306$. All masses are in GeV }  
\label{fig:5}
\efig
%%%%%%%%%%%%%%%%%%%%%%%%%%%%%%%%%%%%%%%%%%%%%%%%%%%%%%%%%%%

%%%%%%%%%%%%%%%%%%%%%%%%%%%%%%%%%%%%%%%%%%%%%%%%%%%%%%%%%%
%%%%%%%%  Section : 4

\section{\label{section:4} \boldmath $\btorholl$ decay mode}

In this section we will calculate the possible measurables associated
with the inclusive decay mode $\btorholl$. Using the defination of the
form factors for $B \to \rho$ transition given by
eqns(\ref{appendixb:eq:1}),(\ref{appendixb:eq:2}),(\ref{appendixb:eq:3})
we can write down the matrix elment as :
\beqa
{\cal M}^{\bar{B} \to \rho} 
  &=& \Bigg[ i \amnab \epsilon^{\nu *} \pcb^\beta q^\beta A ~+~
       \epsilon_\mu^* B ~+~ (\epsilon^*.q) (\pcb)_\mu C \Bigg] ~ (\bl
        \gom \ell)                  \nonumber     \\
  && ~+~ \Bigg[ i \amnab \epsilon^{\nu *} \pcb^\alpha q^\beta D ~+~
       \epsilon_\mu^* E ~+~ (\epsilon^*.q) (\pcb)_\mu F \Bigg] ~ (\bl
        \gom \ell) ~+~  G~(\epsilon^* . q) (\bl ~\ell) ~+~ H~
       (\epsilon^* . q)~ (\bl \gf ~\ell) 
\label{section4:eq:1}
\eeqa
where
\beqa
A &=&  4 ~ \frac{\csev}{s}~ \mb ~T_1(s) ~+~ \cn ~\frac{V(s)}{(\mcb +
\mrh)}           \label{section4:eq:2}                       \\
B &=&  - ~2~ \frac{\csev}{s} ~\mb ~(\mcb^2 - \mrh^2) ~ T_2(s) ~-~ {1
\over 2} (\mcb + \mrh) A_1(s) \cn     \label{section4:eq:3}   \\
C &=&  4 ~ \frac{\csev}{s} \mb \left\{ T_2(s) + \frac{s}{(\mcb^2 -
       \mrh^2)} ~ T_3(\sh) \right\} ~+~ \cn ~ \frac{A_2(s)}{\mcb +
\mrh}                          \label{section4:eq:4}   \\
D &=&  \ct ~ \frac{V(s)}{\mcb + \mrh}   \label{section4:eq:5}  \\
E &=&  - ~ {1 \over 2} (\mcb + \mrh) A_1(s)  \label{section4:eq:6} \\
F &=&  \ct ~ \frac{A_2(s)}{\mcb + \mrh}      \label{section4:eq:7} \\
G &=& - ~ \cqo ~ \frac{\mrh A_0(s)}{\mb}     \label{section4:eq:8} \\
H &=& - ~ \cqt ~ \frac{\mrh A_0(s)}{\mb} ~-~ \ct ~ \frac{m_\ell
        A_2(s)}{\mcb + \mrh} ~+~ \frac{2 \mrh m_\ell}{s} ~ (A_3(s) -
        A_0(s)) ~ \ct
\eeqa

from the above expression of the matrix element we can get the
expression of the partial decay rate :
\beq
\frac{d \Gamma(\bbtorholl)}{d \sh} 
   = \frac{G_F^2 \mcb^5 \alpha^2}{3 \times 2^{10} \pi^5} ~
       |\vtbvtd|^2 ~\lambda^{1/2}(1,\sh,\mrhohs) ~\faco ~\Sigma_\rho  
\label{section4:eq:10}
\eeq
with
\beqa
\Sigma_\rho 
  &=& 
  (1 + \frac{2 \mlhs}{\sh}) ~\lamrho 
     \Bigg[ 4 ~ \mcb^2 ~\sh |A|^2  
     ~+~ \frac{2}{\mcb^2 \mrhohs } 
      ~ (1  + 12 \frac{\mrhohs \sh}{\lamrho}) |B|^2   \nonumber   \\ 
  && +~ \frac{\mcb^2}{2 \mrhohs} ~ \lamrho |C|^2
   +~ \frac{2 }{\mrhohs }  (1 - \mrhohs + \sh) Re(B^* C)  
     \Bigg]
   ~+~ 4 ~\mcb^2 ~\lamrho (\sh - 4 \mlhs) |D|^2     \nonumber   \\ 
  && +~  \frac{2}{\mcb^2}\frac{\Bigg[ 2 (2 \mlhs + \sh) 
     - 2 (2 \mlhs + \sh) (\mrhohs + \sh) + 2 \mlhs (\hat{m}_\rho^4 -
         26 \mrhohs + \sh^2) + \sh ( \hat{m}_\rho^4 + 10 \mrhohs \sh +
           \sh^2) \Bigg]}{\mrhohs \sh} |E|^2      \nonumber   \\
  && +~ \frac{\mcb^2}{2 \mrhohs \sh} \lamrho \Bigg[ (2 \mlhs + \sh)
     (\lamrho + 2 \sh + 2 \mrhohs) - 2 \left\{ 2 \mlhs ( \mrhohs - 5
       \sh) + \sh ( \mrhohs + \sh) \right\} \Bigg] |F|^2  \nonumber \\
  &&  - 3 ~\frac{4 \mlhs - \sh}{\mrhohs} \lamrho |G|^2 
      ~+~ 3 ~\frac{\sh}{\mrhohs} \lamrho |H|^2   
    ~+~ \frac{2 \lamrho}{\mrhohs \sh} \Bigg[ - 2 \mlhs ( \mrhohs - 5
      \sh)  + (2 \mlhs + \sh)               \nonumber  \\
  &&  - \sh (\mrhohs + \sh) \Bigg] Re(E^* F) 
    + \frac{12 \mlh}{\mcb \mrhohs} \lamrho Re(H^* E) 
   + \frac{2 \mcb \mlh}{\mrhohs} \lamrho (1 - \mrhohs + \sh) 
     Re(H^* F)  
\label{section4:eq:11}
\eeqa
%%%%%%%%%%%%%%%%%%%%%%%%%%%%%%%%%%%%%%%%%%%%%%%%%%%%%%%%%%%
%%% Figure : 6
%\vskip 0.7cm
\bfig[t]
\epsfig{file=btorhodrs.eps,height=7.5cm}
\caption{Branching ratio for $\bbtorhott$. Other parameters are :
mSUGRA $m = 200 ~,~ M = 450 ~,~ A = 0 ~,~ \tgb = 35 ~,~ sgn(\mu) $ is
taken to be +ve for rSUGRA $m_A = 306$. All masses are in GeV }
\label{fig:6}
\efig
%%%%%%%%%%%%%%%%%%%%%%%%%%%%%%%%%%%%%%%%%%%%%%%%%%%%%%%%%%

For $B \to \rho$ transition we can find FB asymmetry to be :
\beqa
A_{FB}(\bbtorholl) 
  &=& \left\{- 12 \lambda^{1/2}(1,\sh,\mrhohs) \faco \sh 
      \Bigg[ Re(A^* D) + Re(A^* E) \Bigg] 
      - 3 \frac{\mlh \lambda^{1/2}(1,\sh,\mrhohs)}{\mrhohs}
         \faco            \right.     \nonumber  \\
  &&  \left. \Bigg[ 2 Re(G^* B) \frac{(1 - \mrhohs - \sh)}{\mcb}
         + Re(G^* C) \mcb \lamrho   \Bigg] 
      \right\}/\Sigma_\rho
\label{section4:eq:12}
\eeqa

%%%%%%%%%%%%%%%%%%%%%%%%%%%%%%%%%%%%%%%%%%%%%%%%%%%%%%%%%%%
%%% Figure : 7
\bfig[pb]
\vskip 0.8cm
\epsfig{file=btorhofb.eps,height=7.5cm}
\caption{FB asymmetry for $\bbtorhott$. Other parameters are :
mSUGRA $m = 200 ~,~ M = 450 ~,~ A = 0 ~,~ \tgb = 35 ~,~ sgn(\mu) $ is
taken to be +ve for rSUGRA $m_A = 306$. All masses are in GeV }
\label{fig:7}
\efig
%%%%%%%%%%%%%%%%%%%%%%%%%%%%%%%%%%%%%%%%%%%%%%%%%%%%%%%%%%

We will discuss the CP violation in $B \to \rho$ transition and
discuss the polarization asymmetries associated with final state
lepton in next subsections.

%%%%%%%%%%%%%%%%%%%%%%%%%%%%%%%%%%%%%%%%%
%%%%%  SUBSECTION : 4.1
\subsection{\label{section4:subsec:1} CP violation}

To find out the CP violating partial width asymmetry we requir the
expression of the partial width of $\btobrholl$. The expression of
partial decay rate for $\btobrholl$ is :
\beq
\frac{d \Gamma(\btobrholl)}{d \sh} 
   \ =\ \frac{G_F^2 \mcb^5 \alpha^2}{3 \times 2^{10} \pi^5} ~
       |\vtbvtd|^2 ~\lambda^{1/2}(1,\sh,\mrhohs) ~\faco ~ ( \Sigma_\rho
~+~ 4 \ Im \lamu  ~\bturho ) 
\label{section4:eq:13}
\eeq
with
\beqa
\bturho 
 &=& 
  \Bigg[ Im(\xi_1^* \xi_2) 
    \left\{ 4 \sh \frac{|V(s)|^2}{1 + \mrhohs}  + (1 + \mrhohs) 
       \left( \frac{6 \sh}{\lamrho} + \frac{1}{2 \mrhohs} \right) 
        |A_1(s)|^2  ~+~ \frac{\lamrho}{2 \mrhohs (1 + m_\rho)^2} |A_2(s)|^2 
    \right.                      \nonumber \\ 
 && \left. -~ \frac{1 - \mrhohs - \sh}{\mrhohs} A_1(s) A_2(s) 
   \right\} 
   ~+~ 2 ~ \frac{\csev \hat{m}_b}{\sh} Im (\xi_2) 
    \left\{ 8 \frac{T_1(s) V(s) \sh}{1 + \hat{m}_\rho} 
    \right.                 \nonumber   \\
 && \left. 
      ~+~ 2 A_1(s) T_2(s) (1 + \hat{m}_\rho)^2 (1 - \hat{m}_\rho) 
         \left( 6 \frac{\sh}{\lamrho} + \frac{1}{2 \mrhohs} \right) 
      ~+~ A_2(s) 
           \left( T_2(s) + \frac{\sh}{1 - \mrhohs} T_3(s) \right) 
           \frac{\lamrho}{\mrhohs (1 + \hat{m}_\rho)} 
     \right.                        \nonumber \\ 
 &&  \left.
       - (1 + \hat{m}_\rho) A_1(s) 
           \left(T_2(s) + \frac{\sh}{1 - \mrhohs} T_3(s) \right)
           \frac{1 - \mrhohs - \sh}{\mrhohs}
     \right.                        \nonumber  \\
 &&  \left.
     +~ A_2(s) T_2(s) (1 - \hat{m}_\rho) 
         \frac{1 - \mrhohs - \sh}{\mrhohs}
     \right\}
  \Bigg] (1 + \frac{2 \mlhs}{\sh}) \lamrho
\label{section4:eq:14}
\eeqa
plugging in the expressions of differential decay rate of $\bbtorholl$
and $\btobrholl$ given by eqn(\ref{section4:eq:10}) and
eqn(\ref{section4:eq:13}) respectively we can get the expression of
partial width CP asymmetry :
\beq
A_{CP}(\sh) ~=~ \frac{-2 \ Im \lamu ~ \bturho}{\Sigma_\rho ~+~ 2
~Im\lamu  ~  \bturho} 
\label{section4:eq:15}
\eeq
with $\Sigma_\rho$ and $\bturho$ as given in eqn(\ref{section4:eq:11})
and eqn(\ref{section4:eq:14}).

%%%%%%%%%%%%%%%%%%%%%%%%%%%%%%%%%%%%%%%%%%%%%%%%%%%%%%%%%%%
%%  Figure : 8
%\vskip 1cm
\bfig[t]
\epsfig{file=btorhocpdr.eps,height=8cm,width=12cm}
\caption{CP violating asymmetry ($A_{CP}$) in  $\bbtorhott$ and
$\btobrhott$. The Wolfenstein parameter we have choosen are : $\rho = -
0.07 ~,~ \eta = 0.34 $. Other parameters are : mSUGRA $m = 200 ~,~ M =
450 ~,~ A = 0 ~,~ \tgb = 35 ~,~ sgn(\mu) $ is taken to be +ve for
rSUGRA $m_A = 306$. All masses are in GeV } 
\label{fig:8}
\efig
%%%%%%%%%%%%%%%%%%%%%%%%%%%%%%%%%%%%%%%%%%%%%%%%%%%%%%%%%%%

Another measure of CP violation could be the sum of the FB asymmetries
of $\bbtorholl$ and $\btobrholl$. One can calculate this by use of
eqn(\ref{section4:eq:12}) for $\bbtorholl$ and for $\btobrholl$ by
making appropiate change in the expression of $\cn$. The final
expression is :
\beqa
\delta_{FB} 
  &=&  A_{FB}(\bbtorholl) ~+~ A_{FB}(\btobrholl)  \nonumber   \\
  &=& - ~6 ~ \lambda^{1/2}(1,\sh,\mrhohs) ~ \faco ~ Im \lamu ~  
    \frac{ \left[ Im \xi_2 ~\Gamma_1 ~-~ 2 ~\bturho ~\left\{Re(\cn)
      \Gamma_1 ~+~ \frac{2 \ \csev \ 
       \hat{m}_b}{\sh} ~\Gamma_2 ~\right\} \right]}{\Sigma_\rho ~
       ( \Sigma_\rho ~+~ 4~ Im \lamu ~\bturho)} 
\label{section4:eq:16}
\eeqa
with $\Gamma_1$ and $\Gamma_2$ given as :
\beqa
\Gamma_1 &=& 4 \sh A_1(s) V(s) \ct 
   - \frac{\mlh (1 - \mrhohs - \sh) (1 + \hat{m}_\rho)}{2 \hat{m}_\rho
    \hat{m}_b} A_0(s) A_1(s) \cqo ~+~ \frac{\mlh \lamrho}{\hat{m}_\rho
       \hat{m}_b (1 + \hat{m}_\rho)} A_0(s) A_2(s) \cqo    \\ 
\Gamma_2 &=& 4 \sh T_2(s) V(s) (1 - \hat{m}_\rho) \ct
       ~+~ 4 \sh (1 + \hat{m}_\rho) A_1(s) T_1 \ct 
       ~-~ \frac{(1 - \mrhohs - \sh)(1 - \mrhohs) \mlh}{\hat{m}_\rho
         \hat{m}_b}  A_0(s) T_2(s) \cqo   \nonumber  \\
   &&  \frac{2 \lamrho \mlh}{\hat{m}_\rho \hat{m}_b} A_0(s) 
         \left(T_2(s) + \frac{\sh}{1 - \mrhohs} T_3(s) \right) \cqo 
\label{section4:eq:17} 
\eeqa

%%%%%%%%%%%%%%%%%%%%%%%%%%%%%%%%%%%%%%%%%%%%%%%%%%%%%%%%%%%
%%  Figure : 9
%\vskip 0.6cm
\bfig[t]
\epsfig{file=btorhocpfb.eps,height=8cm,width=12cm}
\caption{CP violating asymmetry ($\delta_{CP}$) in  $\bbtorhott$ and 
$\btobrhott$. The Wolfenstein parameter we have choosen are : $\rho = -
0.07 ~,~ \eta = 0.34 $. Other parameters are : mSUGRA $m = 200 ~,~ M =
450 ~,~ A = 0 ~,~ \tgb = 35 ~,~ sgn(\mu) $ is taken to be +ve for
rSUGRA $m_A = 306$. All masses are in GeV } 
\label{fig:9}
\efig
%%%%%%%%%%%%%%%%%%%%%%%%%%%%%%%%%%%%%%%%%%%%%%%%%%%%%%%%%%%

%%%%%%%%%%%%%%%%%%%%%%%%%%%%%%%%%%%%%%%%%
%%%%%  SUBSECTION : 4.2

\subsection{\label{section4:subsec:2} Polarization asymmetries}
Finally we calculate the three polarization asymmetries namely
longitudinal, transverse and normal, for $\bbtorholl$. 

\par Longitudinal Polarization asymmetry ($P_L$) is : 
\beqa
P_L 
&=& \left\{ 
  24 ~ Re(A^* B) ~(1 - \mrhohs - \sh) \sh \left( - 1 + \faco \right)   
  ~+~ 4 ~\mcb^2 ~\lamrho \sh \faco Re(A^* D)  
    \right.   \nonumber   \\
 && \left.
    +~ \frac{1}{\mrhohs} \left(3 + \faco \right) 
    \Bigg[ 2 ~Re(B^* E) (1 + \hat{m}_\rho^4 + 2 \mrhohs \sh + \sh^2 - 2
        (\mrhohs + \sh))                
    \right.       \nonumber   \\ 
 && \left.
    +~ \mcb^2 Re(C^* E) (1 - 3 (\mrhohs + \sh) 
        - (\mrhohs - \sh)^2 (\mrhohs + \sh) + ( 3 \hat{m}_\rho^4 + 2
         \mrhohs \sh + 3 \sh^2))  \Bigg]
    \right.     \nonumber   \\
 && \left.
    +~ \frac{1}{\mrhohs} \left( Re(B^* F) ( 1 - \mrhohs - \sh) ~+~
      Re(C^* F) \mcb^2 \lamrho \right)
   \Bigg[ \left(3 + \faco \right)~(1 + \mrhohs (\mrhohs - \sh) - 2
       \mrhohs)       
    \right.     \nonumber  \\
 && \left.
   +~ \left(3 - 7 \faco\right) \sh (\mrhohs - \sh) - 8
       \sh \faco \Bigg] 
     \right\}/\Sigma_\rho
\label{section4:eq:18}
\eeqa

%%%%%%%%%%%%%%%%%%%%%%%%%%%%%%%%%%%%%%%%%%%%%%%%%%%%%%%%%%%
%%  Figure : 10
\vskip 0.6cm
\bfig[t]
\epsfig{file=btorholong.eps,height=8cm,width=12cm}
\caption{Longitudinal Polarization asymmetry ($P_L$) in
$\bbtorhott$. The Wolfenstein parameter we have choosen are : $\rho =
- 0.07 ~,~ \eta = 0.34 $. Other parameters are : mSUGRA $m = 200 ~,~ M
= 450 ~,~ A = 0 ~,~ \tgb = 35 ~,~ sgn(\mu) $ is taken to be +ve for 
rSUGRA $m_A = 306$. All masses are in GeV } 
\label{fig:10}
\efig
%%%%%%%%%%%%%%%%%%%%%%%%%%%%%%%%%%%%%%%%%%%%%%%%%%%%%%%%%%%

\par Normal Polarization asymmetry ($P_N$) is :
\beqa
P_N 
  &=&  \lambda^{1/2}(1,\sh,\mrhohs) \sqrt(\sh - 4 \mlhs) \pi
   \Bigg[ 2 Im(E^* F) \frac{1 + \mrhohs - \sh}{\mrhohs} 
     ~+~ 2 Im(A^* E ~+~ B^* D)    \nonumber    \\
 && +~  {1 \over 4 \mrhohs }\left\{2 (1 - \mrhohs - \sh) Im(G^* B)
     + \mcb^2 \lamrho Im(G^* C) \right\} 
   \Bigg]
\label{section4:eq:19}
\eeqa

%%%%%%%%%%%%%%%%%%%%%%%%%%%%%%%%%%%%%%%%%%%%%%%%%%%%%%%%%%%
%%  Figure : 11

\bfig[hb]
\vskip 1cm
\epsfig{file=btorhonorm.eps,height=8cm,width=12cm}
\caption{Normal Polarization asymmetry ($P_N$) in
$\bbtorhott$. The Wolfenstein parameter we have choosen are : $\rho =
- 0.07 ~,~ \eta = 0.34 $. Other parameters are : mSUGRA $m = 200 ~,~ M
= 450 ~,~ A = 0 ~,~ \tgb = 35 ~,~ sgn(\mu) $ is taken to be +ve for 
rSUGRA $m_A = 306$. All masses are in GeV } 
\label{fig:11}
\efig
%%%%%%%%%%%%%%%%%%%%%%%%%%%%%%%%%%%%%%%%%%%%%%%%%%%%%%%%%%%

finally Transverse Polarization asymmetry ($P_T$) is : 
\beqa
P_T 
 &=& \lambda^{1/2}(1,\sh,\mrhohs) ~\sqrt{\sh} ~\mlh ~ \pi  
   \Bigg[ - 4~ Re(A^* B)              \nonumber   \\
 && ~+~ \frac{1}{4 \mrhohs \sh} 
     \left\{ 2~ \left( 2 (1 - \mrhohs - \sh) Re(B^* E) 
     ~+~ \mcb^2 \lamrho  Re(C^* E) \right) \right\} 
   \Bigg] 
\label{section4:eq:20}
\eeqa

%%%%%%%%%%%%%%%%%%%%%%%%%%%%%%%%%%%%%%%%%%%%%%%%%%%%%%%%%%%
%%  Figure : 12
\vskip 0.6cm
\bfig[ht]
\epsfig{file=btorhotrans.eps,height=8cm,width=12cm}
\caption{Transverse Polarization asymmetry ($P_T$) in
$\bbtorhott$. The Wolfenstein parameter we have choosen are : $\rho =
- 0.07 ~,~ \eta = 0.34 $. Other parameters are : mSUGRA $m = 200 ~,~ M
= 450 ~,~ A = 0 ~,~ \tgb = 35 ~,~ sgn(\mu) $ is taken to be +ve for 
rSUGRA $m_A = 306$. All masses are in GeV } 
\label{fig:12}
\efig
%%%%%%%%%%%%%%%%%%%%%%%%%%%%%%%%%%%%%%%%%%%%%%%%%%%%%%%%%%%

%%%%%%%%%%%%%%%%%%%%%%%%%%%%%%%%%%%%%%%%%%%%%%%%%%%%%%%%%%%
%%%%%%%%  Section : 5

\section{\label{section:5} Results and Discussion}

We have performed the numerical analysis of all the kinematical
variables which we have evaluated in Section \ref{section:3} and
Section \ref{section:4}. 

\par For our numerical analysis we would be using the Minimal
Supersymmetric Standard Model (MSSM), which is the simplest (and the
one having the least number of parameters) extension of the SM. 
Actually MSSM itself has fairly large number of parameters which makes
it difficult to do phenomenology with it. We therefore resort to
models which reduces the large parameter space of MSSM to a manageable
level. The models which are around includes minimal Supergravity
(mSUGRA), no-scale, dilaton etc. models. For our analysis
we will be going to use Supergravity (SUGRA) model. The basic features
of all these models is that they assume some sort of unification of
the parameters at some unifying scale. In the numerical analysis of
SUGRA models (both mSUGRA and rSUGRA) the parameters we have choosen
satisfies the radiative electroweak symmetry breaking condition.    

\par In mSUGRA model unification of all the scalar masses, all
gaugino masses and all coupling constants unification is assumed at
GUT scale. So effectively we would be left with five parameters (over
the SM parameters)  at GUT scale they are : $m$ (unified mass of all
the scalars), $M$ (unified mass of all the gauginos), $A$ (unified
trilinear coupling constants), $\tgb$ (the ratio of vacuum expectation
values of the two Higgs doublets) and finally the $sgn(\mu)$
\footnote{our sign of convention of $\mu$ is such that $\mu$ enters
the chargino mass matrix with positive sign}. As emphasized in many
works
\cite{Choudhury:1999ze,RaiChoudhury:1999qb,goto1}
the universality of the scalar masses is not a necessary requirement
and one can relax this. The only constraint for this relaxation if
$K^0 - \bar{K}^0$ mixing. To suppress this mixing it is sufficient to
give unified scalar mass to all the squarks but the Higgs sector can
be given a different unified mass. We will be going to explore sort of
SUGRA model also, which we will be going to call as rSUGRA (or simply
SUGRA) model. In this sort of model we will take mass of the
pseudo-scalar Higgs to be another parameter. For our MSSM parameter 
space analysis we will be going to take 95\% CL bound \cite{expbsg} :
$$
2 \times 10^{-4} ~<~ Br(B \to X_s \gamma) ~<~ 4.5 \times 10^{-4}
$$ 
which is in agreement with CLEO and ALPEH results. As we are more
interested in finding out the effects of NHBs and as well emphasized
in literature that these effects becomes more profound when the final
state leptons are $\tau$ \cite{Grossman:1997qj}. So we will be going
to take the final state leptons to be $\tau$ here. 

\par Our results are summarized in Figures \ref{fig:1} - \ref{fig:12},
where the spikes in the distributions are because of the charm
resonances as given in eqn.(\ref{sec2:eq:1a}).For $\btopitt$, in
Figure(\ref{fig:1}) we have plotted the variation of Branching ratio
with the scaled cm energy of the dileptons. As we can see from the
graphs the deviation from the 
respective SM values is fairly large for almost the whole region of
invariant dilepton mass. The deviation is more profound for SUGRA
model then mSUGRA model. In Figure(\ref{fig:2}) we have plotted the FB
asymmetry for the transition. As has already been noted in earlier
works \cite{Demir:2002a} that in SM the FB asymmetry
vanishes. But if we consider SUSY then one can have a finite value of
FB asymmetry. So any observation of FB asymmetry in this decay mode
($\btopitt$) should be a clear signal of new physics. In
Figure(\ref{fig:3}) we have given the estimates of the CP violating
partial width asymmetry. As expected from the result of
eqn(\ref{section3:eq:15}) the new wilson coefficients doesn't give any
contributions to the numerator of the asymmetry but the denominator
(which essentially is the decay rate) gets contributions from NHBs
and hence the results that NHB effects actually lowers the SM
estimates of the CP violating partial width asymmetry. The reduction
is more for SUGRA model where all the scalar masses are not unified,
in this case one can take Higgs mass to be also a parameter
\footnote{here we have taken pseudo-scalar Higgs mass to be a
parameter and all the rest Higgs masses can be evaluated in terms of
this}. The new Wilson coefficients $\cqo$ and $\cqt$ crucially depends
on Higgs mass and if Higgs mass is low their value is high. In SUGRA
model one can have relatively lower Higgs mass and hence high value of
new Wilsons and hence high partial decay rate which effectively reduces
the CP violating partial width asymmetry. But there exists another
measure of CP violation which is the sum of the FB asymmetries of
$\bbtopill$ and $\btobpill$. This is a sort of measurement which can
be done in a environment having equal numbers of $B$ and $\bar{B}$
pairs and as argued earlier \cite{RaiChoudhury:1997a} doesn't require
any tagging. The important point over here is that FB asymmetry in
this decay is zero in SM and hence the sum of the FB asymmetries of
$\bbtopitt$ and $\btobpitt$. So the parameter $\delta_{FB}$ (which we
have introduced in eqn(\ref{section3:eq:17}) is zero. But if we
consider SUSY then this parameter can have a finite value. In fact as
we have shown in Figure(\ref{fig:4}) $\delta_{FB}$ is more for SUGRA
model then mSUGRA model (which is contrary to the partial width CP
asymmetry). So this quantity could also turn out to be important probe
for new physics. In Figure(\ref{fig:5}) we have plotted the normal
polarization asymmetry of the final state lepton $\ell^-$. As we can
see from the expression of eqn(\ref{section3:eq:21}) the value of
$P_N$ is zero in SM. So the observation of non-zero $P_N$ could also
be interpreted as signal of some new physics \footnote{$P_N$ is a
T-odd observable which comes because of the non-hermiticity of the
effective Hamiltonian, associated with the real $c \bar{c}$
intermediate states, so it can't be taken as a measure of CP
violation}. Rest two polarization asymmetries the longitudinal ($P_L$)
and transverse ($P_T$) vanishes with or without NHBs. 

\par In Figure(\ref{fig:6}) we have plotted the
Branching ratio of $\bbtorhott$ with scaled invariant mass of
dileptons. As we can see there is a fairly large deviation from the SM
value. In Figure(\ref{fig:7}) we have plotted the variation of FB
asymmetry with $\sh$ again one can observe the variation of mSUGRA and
SUGRA results from SM values. Both the partial decay rate and FB
asymmetry increase as compared to SM values in both mSUGRA and SUGRA
models. In Figure(\ref{fig:8}) we have plotted the CP violating
partial width asymmetry. As we can see that here the predictions of
mSUGRA and SUGRA model gets decreased as compared to the SM value. The
reason is the same as explained for $\btopitt$ transition. But again
here if we look at the CP violation through FB asymmetry
(Figure(\ref{fig:9}) and we have the same effect as there in
$\btopitt$ the SUGRA models increases values of as compared to SM
values. Although here the SM values of FB asymmetry as well as
$\delta_{FB}$ is not zero but still SUGRA models gives a enhancement
of more than a order (\ie about a factor of 10) for almost whole
region of the invariant dileptonic mass. In
Figures(\ref{fig:10}),(\ref{fig:11}),(\ref{fig:12}) we have plotted
the Longitudinal ($P_L$), Normal ($P_N$) and Transverse ($P_T$)
polarization respectively respectively. All the three shows variation
from SM values but the general tend is that all these polarization
asymmetries tends to decrease in SUGRA models as compared to SM values
for almost whole region of invariant mass. 
%% New addition
For all the plots we have taken the variation of all the kinematical
variables with the dileptonic invariant mass because the variation
with respect to (wrt) invariant mass has more information than the
result which we get after integrating over the invariant mass and also
the variation of kinematical variables wrt invariant mass are in
principle accessible experimentally. In future B-factories (like
Tevatron and LHCb) more than $10^{11}$, $b \bar{b}$ pairs are expected
to be produced \cite{expbsg}; this is many orders more than the
projected yeild at the $e^+ e^-$ B factories, so these processes can
be observed and some of the measurable quantities in these processes
can be estimated. These processes ($\btopill$ and $\btorholl$) are
useful because they are relatively clean (both theoretically and
experimentally). Also if there is no new source of CP violation
(except CKM phase) then dileptonic decays $\btopill$ and $\btorholl$
should be the first one where CP violation would be observed. As we
can see from the Tables \ref{table:1}, \ref{table:2} that the
kinematical variables of $\btorhott$ looks to be more promising
because of their magnitude. But in $\btopitt$ there are some
distributions like FB-asymmetry, $\delta_{FB}$ and $P_N$ which
vanishes in SM but gives finite (although small) values with
SUSY. This point has already been noted about FB-asymmetry in many
other earlier works \cite{Demir:2002a}. The same phenomena occurs for
$\delta_{FB}$ and $P_N$.  
%%%%%%
%%%%%%% Table
\begin{table}[t]
\caption{\label{table:1} Integrated kinematical variables for
$\btopitt$. The parameters of mSUGRA and rSUGRA are same as given in
Figures(\ref{fig:1}) - (\ref{fig:5})} 
\begin{tabular}{c c c c} \hline
\hspace{.5cm} Variable \hspace{1cm} & \hspace{.5cm} SM \hspace{.5cm} 
& \hspace{.5cm} mSUGRA \hspace{.5cm} & \hspace{.5cm} rSUGRA
\hspace{.5cm} \\ \hline  
$d\Gamma/d\sh \times 10^8$ & 2.6  & 3.43 & 5.04  \\
${\rm A_{FB} \times 10 }$ & 0  & - 0.224  &  - 0.228  \\
${\rm A_{CP}} \times 10^2$ & 0.51  &  0.2 & 0.1  \\
${\rm \delta_{FB}} \times 10^3$  & 0 & - 0.6 & - 0.9 \\
${\rm P_N} \times 10^2$ & 0 & 0.96 & 0.99 \\ \hline
\end{tabular}
\end{table}
%%%%%%%%%%%%%%%%%%

%%%%%%% Table
\begin{table}[t]
\caption{\label{table:2} Integrated kinematical variables for
$\btorhott$. The parameters of mSUGRA and rSUGRA are same as given in
Figures(\ref{fig:6}) - (\ref{fig:12})} 
\begin{tabular}{c c c c} \hline
\hspace{.5cm} Variable \hspace{1cm} & \hspace{.5cm} SM \hspace{.5cm} 
& \hspace{.5cm} mSUGRA \hspace{.5cm} & \hspace{.5cm} rSUGRA
\hspace{.5cm} \\ \hline  
$d\Gamma/d\sh \times 10^8$ & 3.9  & 4.4 & 5.0  \\
${\rm A_{FB} \times 10 }$ & -0.72  & - 0.60  &  - 0.32  \\
${\rm A_{CP}} \times 10$ & 0.13  &  0.09 & 0.04  \\
${\rm \delta_{FB}} $  & 0.0003 & - 0.11 & - 0.18 \\
${\rm P_L}$ & 0.109  & 0.0924  &  0.055  \\
${\rm P_N} \times 10$ & 0.16 & 0.14 & 0.1 \\ 
${\rm P_T}$ & 0.17 & 0.14 & 0.07  \\ \hline
\end{tabular}
\end{table}
%%%%%%%%%%%%%%%%%%

\par Finally our conclusions regarding SUSY effects over a wide range
ot SUSY parameters can be summarized as :
\begin{enumerate}
\item{} {\bf Branching Ratios} ~:~ The branching ratios for both
$\btopitt$ and $\btorhott$ shows large deviation from respective SM
values for almost whole region of the invariant mass.
\item{} {\bf FB asymmetry} ~:~ FB asymmetry for $\btopill$ vanishes
within SM. A non-vanishing FB asymmetry for $\btopill$ clearly gives
indications of some new physics. FB asymmetry for $\btorholl$ shows
significant increase in mSUGRA and rSUGRA models as compared to SM
values. 
\item{} {\bf \boldmath CP violating partial width asymmetry
($A_{CP}$)} ~:~  The predictions of mSUGRA and SUGRA models is to
reduce this asymmetry for both $\btopitt$ and $\btorhott$ as compared
to SM values. 
\item{} {\bf \boldmath CP violation from FB asymmetry ($\delta_{FB}$)}
~:~  This observable vanishes in SM for $\btopitt$ a non-vanishing
value of this clearly indicates new physics effects. For $\btorhott$
the SM value prediction is very low, both SUGRA and mSUGRA can give
enhancement of over a order almost for whole region of scaled dilepton
invariant mass. 
\item{} {\bf Polarization asymmetries} ~:~ For $\btopitt$ all the
three polarization asymmetries (longitudinal, normal and transverse)
vanishes in SM. If we include NHB effects then although longitudinal
and transverse still remains zero the normal polarization asymmetry
becomes non-zero. So observation of normal polarization asymmetry can
still be regarded as evidence for new physics. For $\btorholl$ all the
three polarization asymmetries decreases, with respect to their
respective SM values, by switching on the NHB effects. 
\end{enumerate}

The observation of the decay modes $\btopill$ and $\btorholl$ can be
expected to be a very useful tool for the search of new physics
effects as well as for the measurement of the CP violating parameters
of CKM matrix.

%%%%%%%%%%%%%%%%%%%%%%%%%%%%%%%%%%%%%%%%%%%%%%%%%%%%%%%%%%%%%%

\begin{acknowledgments}
This work was supported under SERC scheme of Department of Science and
Technology (DST), India. 
\end{acknowledgments}

%%%%%%%%%%%%%%%%%%%%%%%%%%%%%%%%%%%%%%%%%%%%%%%%%%%%%%%%%%%
%%%%%%%%%    APPENDIX

\appendix

%%%%%%%%%%%%%%%%%%%%%%%%%%%%%%%%%%
%%  Appendix : cons

\section{\label{appen:con} Input Parameters}

\bcen 

$ m_u ~=~ m_d ~=~$ 10 MeV   \\
$\mb ~=~ 4.8 $ GeV  \ , \ $ m_c ~=~ 1.4$ GeV  \,\ $m_t ~=~ 176$ GeV \\
$\mcb ~=~ 5.26$ GeV \ , \ $\mpi ~=~ 0.135$ GeV \ , \ $m_\rho ~=~ 0.768
$ GeV \\   
$|\vtbvtd| ~=~$ 0.011 \ , \ $\alpha ~=~ \frac{1}{129} \ , \ G_F ~=~
{\rm 1.17 \times 10^{-5} ~ GeV^{-2} }$  \\ 
$m_\tau ~=~ $ 1.77 GeV \ , \ $\tau_B ~=~ {\rm 1.54 \times 10^{-12}}$ s
\\
Wolfentein parameters : $ \rho ~=~ - 0.07 \ , \ \eta ~=~ 0.34$

\ecen

%%%%%%%%%%%%%%%%%%%%%%%%%%%%%%%%%%
%%  Appendix : A

\section{\label{appen:a} Form factors for $B \to \pi$ transition } 

 We will be going to use the form factors given by
Coleangelo \etal \cite{Colangelo:1996} :
\beqa
\la \pi (\ppi) | \bd \gm P_{L,R} b | B (\pcb) \ra &=& {1 \over 2}
 \left\{ (2 \pcb - q )_\mu F_1(\qs) ~+~ \frac{\mcb^2 - \mpi^2}{\qs}
\qm ( F_0(\qs) - F_1(\qs) )  \right\}  
\label{appendixa:eq:1}      \\
\la \pi (\ppi) | \bd i \sigmn \qon P_{L,R} b | B (\pcb) \ra &=& 
{1 \over 2} \left\{ (2 \pcb - q )_\mu - ( \mcb^2 - \mpi^2) \qm
\right\} \frac{F_T(\qs)}{\mcb + \mpi} 
\label{appendixa:eq:2}
\eeqa
To get the matrix element for scalar current we multiply
eqn.(\ref{appendixa:eq:1}) by $\qm$ we get 
\beq
\la \pi (\ppi) | \bd \pr b | B (\pcb) \ra ~=~ \frac{1}{2 \mb} (\mcb^2
- \mpi^2) F_0(\qs)
\label{appendixa:eq:3}
\eeq
where we have neglected the mass of d-quark.

\par The defination of the Form factors $F_0, ~ F_1$ and $F_T$
are \footnote{All the three $F_0 ~,~F_1$ and $F_T$ are not
independent. $F_T$ can be related to $F_0$ and $F_1$ by equation of
motion and the relationship works out to be $F_T ~=~ (\mcb + \mpi) \mb
\frac{F_0 - F_1}{\qs}$} ($\qs$ in the units of ${\rm GeV}^2$) :  
\beqa 
F_0(\qs) &=&  \frac{F_0(0)}{1 - \qs /7^2}    \nonumber  \\
F_1(\qs) &=&  \frac{F_1(0)}{1 - \qs /5.3^2}  \nonumber  \\
F_T(\qs) &=&  \frac{F_T(0)}{ \left(1 - \qs /7^2 \right) 
 ~\left(1 - \qs /5.3^2\right) }        \nonumber    \\
\ftt  &=&  \frac{\ft}{(\mcb + \mpi)} \mb 
\label{appendixa:eq:4}
\eeqa
with $F_0(0) ~=~ 0 ~,~ F_1(0) ~=~ 0.25$ and $F_T(0) ~=~ - 0.14$

%%%%%%%%%%%%%%%%%%%%%%%%%%%%%%%%%%
%%  Appendix B 

\section{\label{appen:b} Form factors for $B \to \rho$ transition}

For $B \to \rho$ transition we will be going to use the form factors
given by  Colangelo \etal \cite{Colangelo:1996} :

\beqa
\la \rho (\prh) | \bd \gm \pl b | \bar{B} (\pcb) \ra 
  &=&  i \amnab \epsilon^{\nu *} \pcb^\alpha q^\beta
    \frac{V(\qs)}{\mcb + \mrh}  - {1 \over 2}~
    \left\{ \epsilon_\mu ( \mcb + \mrh ) A_1( \qs ) 
    \right.                        \nonumber   \\
 && \left. - (\epsilon^* . q)
    (2 \pcb - q)_\mu \frac{A_2(\qs)}{\mcb + \mrh} - \frac{2 \mrh}{\qs}
     (\epsilon^* .q ) \left[A_3(\qs) - A_0(\qs)\right] 
    \right\}                        
\label{appendixb:eq:1}                     \\
\la \rho(\prh) | \bd i \sigmn \qon P_{R,L} b | \bar{B}(\pcb) \ra 
 &=&  - 2 i \amnab \epsilon^{\nu *} \pcb^\alpha q^\beta T_1(\qs) 
     \pm \left[ \epsilon_\mu^* (\mcb^2 - \mrh^2) - (\epsilon^* . q)~(2
      ~ \pcb - q)_\mu \right] T_2(\qs)      \nonumber   \\
 &&  \pm (\epsilon^* . q) \left[ q_\mu - \frac{\qs}{\mcb^2 -
      \mrh^2}~(2 \pcb - q)_\mu \right] T_3(\qs) 
\label{appendixb:eq:2}
\eeqa
where $A_3$ can be written in terms of $A_1$ and $A_2$, \ie
\beq
A_3(\qs) ~=~ \frac{\mcb + \mrh}{2 \mrh} A_1(\qs) - \frac{\mcb -
       \mrh}{2 \mrh} A_2(\qs) 
\label{appendixb:eq:3}
\eeq
In above equations $\epsilon$ is the polarization vector of $\rho$ and
$q = \pcb - p_\rho$ is the momentum transfer. 

\par To get the matrix elment for scalar (or pseudoscalar) current we
multiply both sides of eqn.(\ref{appendixb:eq:1}) by $q^\mu$. On 
simplifying we get :
\beq
\la \rho(\prh) | \bd \pr b | B (\pcb) \ra ~=~ -
\frac{\mrh}{\mb}~(\epsilon^* . q) A_0(\qs)
\label{appendixb:eq:4}
\eeq
The defination of the form factors is \footnote{again here also $T_3$
can be related to $A_3$ and $A_0$ by EOM and relationship is : $T_3
~=~ \mcb \mb \frac{A_3 - A_0}{\qs}$} ($\qs$ in units of ${\rm GeV}^2)$
:  
\beqa
V(\qs) &=& \frac{V(0)}{1 ~-~ \qs /5^2}    \nonumber  \\
A_1(\qs) &=& A_1(0) ~(~1 ~-~ 0.023~ \qs)           \nonumber  \\
A_2(\qs) &=& A_2(0) ~(1 ~+~ 0.034 ~\qs)           \nonumber  \\
A_0(\qs) &=& \frac{A_3(0)}{1 ~-~ \qs /4.8^2}  \nonumber  \\
T_1(\qs) &=& \frac{T_1(0)}{1 ~-~ \qs /5.3^2}  \nonumber  \\
T_2(\qs) &=& T_2(0)~(1 ~-~ 0.02~ \qs)             \nonumber  \\
T_3(\qs) &=& T_3(0)~(1 ~+~ 0.005 ~\qs) 
\label{appendixb:eq:5}
\eeqa
with $V(0) ~=~ 0.47 ~,~ A_1(0) ~=~ 0.37 ~,~ A_2(0) ~=~ 0.4 ~,~ A_0(0)
~=~ 0.3 ~,~ T_1(0) ~=~ 0.19 ~,~ T_2(0) ~=~ 0.19 ~,~ T_3(0) ~=~ -
0.7$.

%%%%%%%%%%%%%%%%%%%%%%%%%%%%%%%%%%%%%%%%%%%%%%%%%%%%%
%  

\end{document}